\documentclass{aastex}
\usepackage{emulateapj5}
\usepackage{times,mathptmx}
\usepackage{graphicx}

\newcommand*{\Msun}{\ensuremath{\mathrm{M_\odot}}}%
\newcommand*{\Mpc}{\ensuremath{\mathrm{Mpc}}}%
\slugcomment{ApJL, 621, 2005 March 10, Accepted 2005 February 1}

\shorttitle{The Specific Star Formation Rate}
\shortauthors{Bauer et al.}

\begin{document}


\title{Specific Star Formation Rates to Redshift 1.5}


\author{A. E. Bauer\altaffilmark{1}, N. Drory\altaffilmark{1},
    G. J. Hill\altaffilmark{1}, and G. Feulner\altaffilmark{2} }

\altaffiltext{1}{University of Texas at Austin, Austin, Texas 78712,
  email: \{amanda,drory,hill\}@astro.as.utexas.edu}
\altaffiltext{2}{Universit\"ats-Sternwarte M\"unchen,
Scheinerstra\ss e 1, D-81679 M\"unchen, Germany, email:
  feulner@usm.uni-muenchen.de}

\begin{abstract}
We present a study to determine how star formation contributes to
galaxy growth since $z=1.5$ over five decades in galaxy stellar mass.
We investigate the specific star formation rate (SSFR; star formation
rate [SFR] per unit galaxy stellar mass) as a function of galaxy
stellar mass and redshift.  A sample of 175 $K$-band selected galaxies
from the MUnich Near-Infrared Cluster Survey spectroscopic dataset
provide intermediate to high mass galaxies (mostly $M_* \geq 10^{10}
\Msun$) to $z=1$.  The FORS Deep Field provides 168 low mass galaxies
(mostly $M_* \leq 10^{10}\Msun$) to $z=1.5$.  We use a Sloan Digital
Sky Survey galaxy sample to test the compatibility of our results with
data drawn from a larger volume.  We find that at all redshifts, the
SSFR decreases with increasing galaxy stellar mass suggesting that
star formation contributes more to the growth of low mass galaxies
than to the growth of high mass galaxies, and that high mass galaxies
formed the bulk of their stellar content before $z=1$.  At each epoch
we find a ridge in SSFR versus stellar mass that is parallel to lines
of constant SFR and evolves independently of galaxy stellar mass to a
particular turnover mass.  Galaxies above this turnover mass show a
sharp decrease in the SFR compared to the average at each epoch and
the turnover mass increases with redshift.  The SFR along the SSFR
ridge decreases by roughly a factor of 10, from 10 $\Msun$yr$^{-1}$ at
$z=1.5$ to 1 $\Msun$yr$^{-1}$ at $z=0$.  High mass galaxies could
sustain the observed rates of star formation over the 10 Gyr observed,
but low mass galaxies likely undergo episodic starbursts.
     
\end{abstract}



\section{Introduction}
In an effort to study galaxy assembly, we look at the contribution of
star formation to the growth of stellar mass in galaxies as a function
of time.  Several groups have determined stellar masses of galaxies
from redshift surveys with multiwavelength observations in the local
universe (2dF, \citealp{Cole01}; SDSS \citealp{Kauffmannetal03a}; SDSS
\& 2MASS, \citealp{Belletal03, MUNICS-SDSS}) and at high redshift
(\citealp{BE00,MUNICS3,Cohen02,Rudnicketal03,DPFB03,Fontanaetal04,MUNICS6}).

Here we investigate the specific star formation rate (SSFR), which
measures the star formation rate (SFR) per unit galaxy stellar mass, to
study explicitly how star formation contributes to galaxy growth for
galaxies of different masses at different times in the history of the
universe.  

The SSFR has been studied at low redshifts
\citep{PerezG03,BCWTKHB04} and intermediate redshifts
\citep{CSHC96,Guzman97,BE00,Fontanaetal03} but no significant study of
SSFR evolution over a wide range of galaxy masses and redshift has
been undertaken.  High mass galaxies at high redshifts are just now
beginning to be studied \citep{Juneau04}.  \citet{CSHC96} used
rest-frame $K$-band (2.2$\mu$m) luminosities and [OII]$\lambda3727$
equivalent widths to show that galaxies with rapid star formation
decrease in $K$ luminosity, and therefore mass, with decreasing
redshift.  \citet{BE00} inferred stellar masses of galaxies and
pointed out a generally increasing SSFR with redshift and a trend for
low mass galaxies to exhibit larger SSFRs.

We combine two complementary redshift surveys to broaden the mass and
redshift range that we can probe.  The wide-area, medium deep MUNICS
\citep{MUNICS1,MUNICS5} spectroscopic dataset provides intermediate to
high mass galaxies typically in the mass range of $M_* \geq
10^{10}\Msun$.  The FORS Deep Field \citep{FDF1,FDF2} covers a small
portion of the sky very deeply, contributing $M_*~<~10^{10}$~\Msun~\ 
galaxies to the sample.

We discuss the samples used and the methodology for determining SFRs
and galaxy masses in Section 2.  We describe the results of this
study, possible selection effects and complete a comparison to the
local universe in Section 3.  In Section 4 we present a discussion of
the physical implications of our results as well as a comparison to
the literature.

Throughout this Letter we adopt an $\Omega_M = 0.3$, $\Omega_{\Lambda} =
0.7$, $H_0 = 72\ \mathrm{km\ s^{-1}\Mpc^{-1}}$ cosmology.  


\section{Galaxy Data}

The MUNICS project is a wide-area, medium-deep, photometric and
spectroscopic survey selected in the $K$-band and reaching
$K\sim19.5$.  It covers nearly one square degree in the $K$ and $J$
bands with follow-up imaging in the $I, R, V,$ and $B$ bands over 0.5
square degrees \citep{MUNICS1}.  Spectroscopy is complete to
$K\sim17.5$ over 0.25 square degrees and reaches $K=19.5$ for 100
square arcmins.  The spectra cover a wide wavelength range of
$4000-8500\mathrm{\AA}$ at $13.2\mathrm{\AA}$ (FWHM) resolution,
sampling galaxies in the redshift range of $0.07<z<1$ \citep{MUNICS5}.
Our MUNICS sample contains 175 objects, which are mostly massive
($M_*> 10^{10}$\Msun) field galaxies with detectable
[OII]$\lambda$3727 emission.

\begin{figure*}[t]
\includegraphics[scale=0.7, width=0.45\textwidth, angle=-90]{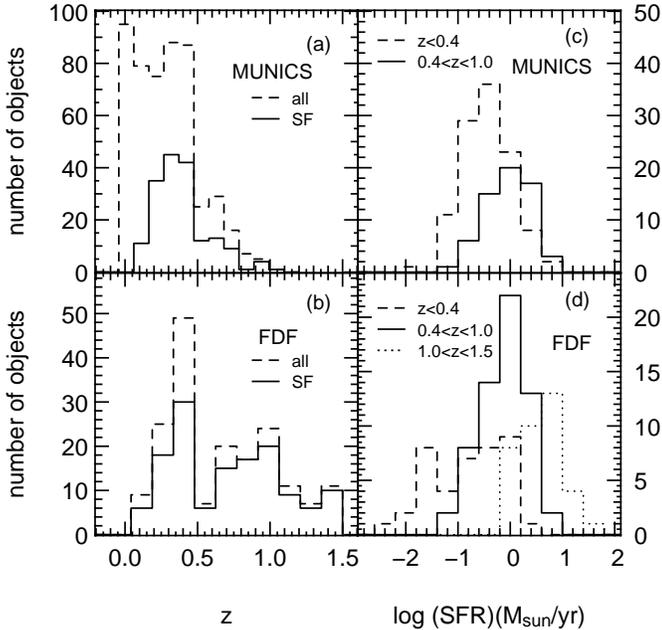}
\caption{\label{f:hist} Distribution of Redshift for MUNICS and FDF
samples.  In panels \textit{a} and \textit{b}, the dashed lines
show the full spectroscopic samples, and the solid lines show jsut the
star-froming galaxies used for this study.  Except at the lowest
redshift, the majority of galaxies have detectable star formation to
the limits of these surveys.  Panels \textit{c} and \textit{d}
show the distribution of SFRs for MUNICS (\textit{c}) and FDF
(\textit{d}) galaxies binned by redshift.  The dashed histograms
represent $z<0.4$, the solid $0.4<z<1.0$ and in panel \textit{d}the
dotted distribution shows galaxies with $1.0<z<1.5$.  The mean and the
maximum SFR in each redshift bin increases with redshift.}
\end{figure*}

The FORS Deep Field (FDF) spectroscopic survey provides low-resolution
spectra with detectable [OII]$\lambda3727$ in the spectral window
($3300-10000\mathrm{\AA}$ at $23\mathrm{\AA}$ (FWHM) resolution) to
$z=1.5$ \citep{FDF2}.  The FDF survey is $I$-band selected reaching
$I_{AB} = 26.8$ with spectroscopy to $I_{AB}=24$.  The FDF covers $7'$
x $7'$ in eight bands: UBgRIzJK \citep{FDF1}.  Our FDF sample includes 168
galaxies with detectable star formation, and masses of mostly $M_*<
10^{10}$\Msun.

Stellar masses of galaxies are determined, as described in
\citet{MUNICS6}, by fitting a grid of composite stellar population
models of varying age, star formation history, and dust extinction to
multi-wavelength photometry to determine mass-to-light ($M/L$) ratios.
The total systematic uncertainty in the $M/L$ ratio is about 25\%
\citet{MUNICS6}.

We have developed a program to automatically measure emission line
fluxes, equivalent widths and continuum breaks from flux calibrated
spectra.  To each rest-frame emission line region of interest, we fit
a Gaussian profile plus a polynomial continuum, using the spectral
resolution as a first guess at the line width.  Errors in measurements
are determined by fitting spectral regions multiple times with
different continuum estimates.  This program gives us a consistent
flux measurement between the galaxy samples measured in this study.

As an SFR indicator, we use the flux of the [OII]$\lambda3727$
emission feature which remains in the spectral window from low
redshift to $z=1.5$.  We use the Kennicutt (1998, Equation 3)
conversion from [OII] line luminosity to SFR in units of solar masses
per year.


\section{Specific Star Formation Rates}

Fig.~\ref{f:hist} demonstrates the similarity between the redshift
distributions of the full spectroscopic samples (dashed lines) and
star forming sub-samples (solid lines) used in this study from MUNICS
(Fig. 1\textit{a}) and FDF (Fig. 1\textit{b}).  The majority of
galaxies in the full samples have detectable star formation and the
distributions match very well, indicating little redshift bias in
making this selection of objects.  There are fewer star forming
galaxies in the MUNICS sample at the lowest redshifts with only six
low redshift MUNICS galaxies excluded because of the
[OII]$\lambda$3727 emission line being blueward of the spectral
coverage.  For MUNICS, 52\% of the spectroscopic sample shows star
formation.

For the FDF galaxy sample, 75\% of all galaxies with $z<1.5$ show
 detectable star formation via the [OII]$\lambda3727$ emission
 feature.  The higher fraction of star forming galaxies in FDF is
 understandable since the FDF spectroscopic absolute magnitude limit
 is deeper than MUNICS and fainter galaxies tend to have more star
 formation (e.g. \citealp{Kenn84}).
 
Fig.~\ref{f:hist} also shows the distribution of SFRs seen among the
two samples of galaxies split into redshift bins.  MUNICS galaxies are
shown in Figure 1\textit{c} with the dashed histogram representing
$z<0.4$ and the solid histogram showing the range of $0.4<z<1.0$.
Figure 1\textit{d} shows the FDF SFR distribution with the same
redshift bins in addition to a high redshift bin shown by the dotted
histogram for $1.0<z<1.5$.  The maximum SFR increases with redshift,
and is consistent between the two samples.

\begin{figure*}[t]
\includegraphics[angle=-90,scale=0.75]{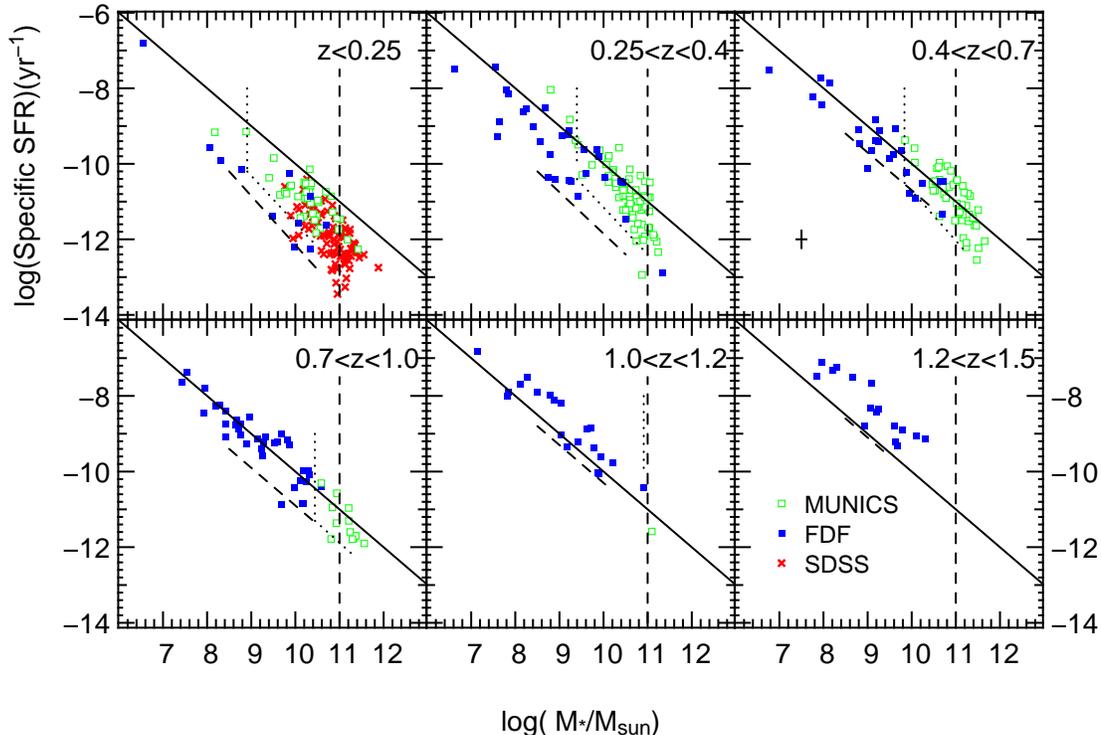}
\caption{\label{f:ssfr} Specific star formation rate versus galaxy
  stellar mass as a function of redshift.  The open squares are MUNICS
  galaxies, the solid squares represent FDF galaxies, and the crosses
  correspond to SDSS galaxies, chosen to match the photometric limits
  of the MUNICS sample.  The solid diagonal line in each redshift bin
  shows a SFR = 1 \Msun yr$^{-1}$ and the vertical dashed line
  references a galaxy with $M_*$ = $10^{11}$\Msun.  Estimated error
  bars, appropriate for a majority of the galaxies, are shown in the
  bottom left corner of the $0.4<z<0.7$ bin.  The vertical dotted line
  indicates the MUNICS mass limit and the diagonal dotted line shows
  the lower limit for MUNICS detectable star formation in each bin.
  The diagonal dashed lines show lower limits for detectable SSFR for
  FDF.}
\end{figure*}

Fig.~\ref{f:ssfr} shows the SSFR versus galaxy stellar mass, as a
function of redshift.  A sample of SDSS galaxies is shown in the
lowest redshift bin as crosses and described in detail below.  The
solid diagonal line in each panel indicates a constant SFR = 1\Msun
yr$^{-1}$.  Lines of constant SFR are parallel to this line and SFRs
increase to the upper right.  The vertical dashed line designates a
galaxy with $M_* = 10^{11}$\Msun.  The short vertical dotted line
shows the limiting MUNICS galaxy stellar mass for each redshift bin.
The diagonal dotted and dashed lines show limits for detectable SSFR
for MUNICS and FDF, respectively, derived from their spectroscopic
sensitivity limits.

Several features are evident at each epoch in this plot.  Low to
intermediate mass galaxies form a ``ridge'' of SSFR that lies parallel
to the solid diagonal line of constant SFR.  The ridge remains as mass
increases until, at some high mass, there appears to be a turnover
mass, above which the SSFR decreases.  This turnover mass increases with
redshift accompanied by higher mass galaxies showing higher SSFRs at
earlier times.  This behavior has been termed 'downsizing' by
\citet{CSHC96}.

There exists a general correlation between SSFR and mass at low
redshift (e.g. \citealp{BE00,BCWTKHB04}.  We show evidence here for a
ridge in SSFR that evolves with redshift independent of galaxy stellar
mass for the full range of $M_* = 10^{7} - 10^{11} \Msun$.  The
average SSFR at a given mass on the ridge decreases from $z=1.5$ to
$z=0$ so that the average SSFR for a low mass galaxy at $z=1.5$ is
roughly 1 dex higher than the average SSFR of a low mass galaxy at
$z=0.25$.  Galaxies with SFR$< 1$ \Msun yr$^{-1}$, below the SSFR
ridge, are increasingly too faint to detect as redshift increases, as
shown by the diagonal dashed line in Fig.~\ref{f:ssfr}.  The region
below the SSFR ridge of galaxies in Fig.~\ref{f:ssfr} could also be
populated by galaxies that are not forming stars at the time of
observation.  Galaxies of $M_* > 10^{11} \Msun$ exist at
low SSFR, showing that they form fewer stars compared to the lower mass
galaxies.  The turnover mass refers to the mass at which the SSFR
noticeably decreases from the ridge.  We note that, while detecting
the lowest values of SFR is affected by incompleteness, more than 50\%
of the galaxies have detectable star formation and the upper bound of
the SSFR ridge has no selection effects.

We make no corrections for extinction to the emission line fluxes in
the present study for multiple reasons.  We only measure H$\alpha$ for
those galaxies at $z<0.3$, and can only measure a Balmer decrement at
low redshift.  In addition, we only have H$\beta$ line flux
measurements for 10\% of the galaxies.  For those galaxies with
measurements of both H$\alpha$ and H$\beta$ we find an average $E(B-V)
= 0.2$, assuming a case B recombination.  Galaxy samples selected at
mid and far infrared wavelengths would likely yield some starbursting
galaxies with higher SFRs and would be affected by dust extinction.
We feel it would be irrelevant to apply a flat, generic low redshift
correction to all galaxies here, because while it would increase the
global star formation density by 50\%, it would not change the
differential star formation effects studied here.  If we were to apply
a mass-dependent extinction correction, the slope of the SSFR ridge
would change, but the differential effects with redshift would remain
identical.

The trends seen among the FDF and MUNICS galaxy surveys exhibit
similar evolution in their overlapping mass range around
$M_*~=~10^{10}$\Msun\ and are therefore suitable to analyze
simultaneously to cover such a large range of galaxy stellar mass.

There are few FDF galaxies present in the lowest redshift bin in
Fig.~\ref{f:ssfr} as anticipated since the FDF covers little volume at
low redshift.  The lack of massive galaxies in the $z>1$ bins also
shows the small volume of the FDF in addition to the spectroscopic
limit for detecting massive galaxies.  In order to verify that the
increase in SSFR at higher redshifts is not due to rarer objects being
seen as larger volumes are probed, we investigate the trends in SSFR
from the SDSS.  We gathered SDSS galaxies \citep{SDSS-DR2} selected to
match the MUNICS magnitude limits from the sample described in
\citet{MUNICS-SDSS}, who determined galaxy stellar masses following
the same method as the rest of our sample.  We used the
[OII]$\lambda$3727 flux as reported by SDSS and followed the
\citet{Kenn98} conversion to SFR.

We chose a random sample of SDSS galaxies to match the number of
galaxies in the lowest redshift bin from MUNICS and FDF and show them
in Fig.~\ref{f:ssfr} as crosses.  The SSFRs of the SDSS galaxies are
consistent with the MUNICS values in this redshift bin, which we
confirmed by repeating the selection multiple times.  Out of the whole
SDSS sample, only 1.15\% lie above the 1 \Msun yr$^{-1}$\ constant SFR
line.  Hence, if the abundance of galaxies above this line at high
redshift were only due to the larger volume sampled, we would expect
only such a small fraction of objects to lie above this line.


\section{Discussion and Conclusions}

We present a study of the contribution of star formation to galaxy
growth from $z=1.5$ to the present.  We investigate the SSFR as a
function of redshift and galaxy stellar mass over five decades in
galaxy stellar mass.  At all redshifts, the SSFR decreases as stellar
mass increases.  This indicates a higher contribution of star
formation to the growth of low mass galaxies since $z=1.5$ and
suggests that high mass galaxies formed the bulk of their stellar
content earlier than $z=1$.  The drop in SSFR for high mass ($M_* =
10^{11}$\Msun) galaxies above the turnover mass for each epoch seen in
Fig.~\ref{f:ssfr} is symptomatic of higher mass galaxies tending to be
the early-type, redder population which forms few stars after $z=1$.
This result is compatible with the detection of massive galaxies at
$z>1$ \citep{Saraccoetal03,Fontanaetal04,GDDSnature04} and the
mass-dependent SFR from $z=0$ to $z=2$ suggested by \citet{Heavens04}.

Fig.~\ref{f:ssfr} shows evidence of a ridge in SSFR that runs parallel
to lines of constant SFR.  The ridge exists for all galaxy stellar
masses $M_*=10^{7} - 10^{11}$\Msun\ and increases uniformly,
independent of mass as redshift increases.  The first evidence for
such a ridge in SSFR was noted by \citet{BE00}.  Our work moves beyond
that study with a mass limited sample at each redshift bin and a wide
range of masses.

The ridge in SSFR shifts downward as redshift decreases indicating a
steady decrease in the global SFR by a factor of 10 from $z=1$ to
$z=0$, as widely noted in the literature (e.g.
\citealp{Madauetal96,CFRS96,CFRS1297,RRetal97,Floresetal99,Tresseetal02}).
Previous studies have not differentiated whether the decline in the
global SFR is experienced by galaxies of all types or specific to a
limited galaxy population.  Our study shows a uniform change in SFR
independent of galaxy stellar mass. 

Using nearly 10$^5$ SDSS galaxies, \citet{BCWTKHB04} study several
properties of star forming galaxies at $z<0.2$.  They show the
observed likelihood distribution of SSFR versus stellar mass in their
Figure 24 which demonstrates the existence of a SSFR ridge.  This
ridge is compatible with the one shown in this work if identical SFR
conversions and extinction corrections are applied to the
[OII]$\lambda$3727 emission line flux as shown in the low redshift
panel of Fig.~\ref{f:ssfr}.  Our comparison to SDSS galaxies reveals
that our findings are not the result of more extreme galaxies being
detected as more volume is sampled.

Galaxies exist on Fig.~\ref{f:ssfr} only when the star formation
induces detectable amounts of [OII]$\lambda$3727 emission.  At any
epoch a majority of galaxies ($50-75\%$) show detectable star
formation with SFRs lying on the SSFR ridge.  Galaxies with $M_* >
10^{10}$\Msun\ could sustain the observed SFRs for a doubling time of
order the Hubble time, without moving significantly on
Fig.~\ref{f:ssfr}, but the SFR is observed to decrease over time,
presumably because of gas depletion.

Galaxies with $M_* < 10^{10}$\Msun\ likely evolve differently.  Low
mass galaxies sustaining the observed SFRs would significantly evolve
on Fig.~\ref{f:ssfr} from $z=1$ to $z=0$ so that we would see more
intermediate mass galaxies with SFR\ $> 1$ \Msun yr$^{-1}$ at $z=0$,
which is not evident.  A galaxy of mass, $M_* = 10^{9}$\Msun, with
constant SFR = 1 \Msun yr$^{-1}$ has a doubling time of the stellar
population of $10^{9}$yrs.  Such a galaxy, with a typical gas fraction
of order 1 \citep{Bothun84,Kannappan04}, could only sustain this rate
of star formation for a short period of time before the gas supply
diminishes (if we assume no significant gas infall).  Therefore, it is
likely that these galaxies exist on Fig.~\ref{f:ssfr} during periodic
starbursts.  If gas depletes during each burst then subsequent bursts
would exhibit progressively lower SFRs, consistent with the observed
trend.  The duty cycle must be at least 50\%, since the majority of
the galaxies have detected star formation.

If there exists a transition mass between the two separate modes of
star formation (sustained or periodic bursts), then the coherence of the
trend of the SSFR ridge decreasing with redshift independent of mass
is particularly interesting.

\acknowledgments

We wish to thank Karl Gebhardt and Sheila Kannappan for useful
discussions.  Niv Drory acknowledges support from the Alexander von
Humboldt Foundation.  We thank the MUNICS and FDF collaborations for
making their data available.  This research was partly supported by
NSF grant 9987349.


\end{document}